\documentclass[amsmath,amssymb,aps,pra,twocolumn]{revtex4}

\usepackage{graphicx}
\usepackage{mathrsfs}

\newcommand{\annihilate}[1]{\hat{#1}}
\newcommand{\create}[1]{\hat{#1}^{\dag}}
\newcommand{\ket}[1]{{|#1\rangle}}

\newcommand{\opa}{\annihilate{a}}
\newcommand{\opad}{\create{a}}
\newcommand{\opb}{\annihilate{b}}
\newcommand{\opbd}{\create{b}}
\newcommand{\opd}{\annihilate{d}}
\newcommand{\opdd}{\create{d}}

\begin{document}

\title{Mechanical resonators for storage and transfer of electrical
  and optical quantum states} \author{S. A. McGee$^1$,
  D. Meiser$^{1,2}$, C. A. Regal$^1$, K. W. Lehnert$^{1,3}$, and
  M. J. Holland$^1$} \affiliation{ $^1$JILA and Department of Physics,
  University of Colorado, Boulder
  CO 80309-0440, USA.\\
  $^2$Tech-X Corp., 5621 Arapahoe Ave. Ste. A, Boulder, CO 80303,
  USA.\\
  $^3$National Institute of Standards and Technology, Boulder CO
  80309,
  USA.\\
} \date{\today}

\begin{abstract}
  We study an optomechanical system in which a microwave field and an
  optical field are coupled to a common mechanical resonator.  We
  explore methods that use these mechanical resonators to store
  quantum mechanical states and to transduce states between the
  electromagnetic resonators from the perspective of the effect of
  mechanical decoherence.  Besides being of fundamental interest, this
  coherent quantum state transfer could have important practical
  implications in the field of quantum information science, as it
  potentially allows one to overcome intrinsic limitations of both
  microwave and optical platforms.  We discuss several state transfer
  protocols and study their transfer fidelity using a fully quantum
  mechanical model that utilizes quantum state-diffusion techniques.
  This work demonstrates that mechanical decoherence should not be an
  insurmountable obstacle in realizing high fidelity storage and
  transduction. 
\end{abstract}

\maketitle

%%%%%%%%%%%%%%%%%%%%%%%%%%%%%%%%%%%%%%%%%%%%%%%%%%%%%%%%%%%%%%%%%%%%%%%%%%%%%%%%
\section{Introduction}

\begin{figure} \includegraphics[width=60mm]{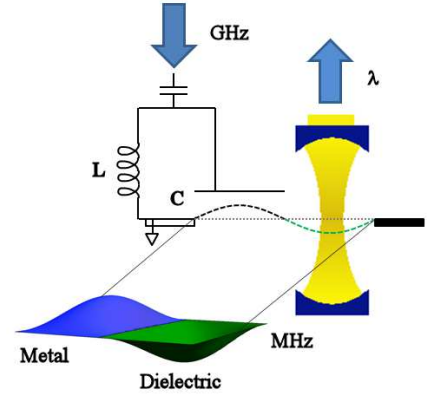}
% \begin{figure} \includegraphics[width=60mm]{pics/EOMDrumCircuitDiagram.eps}
  \caption{(Color online) A schematic diagram for an optical cavity
    coupled to a thin dielectric membrane mechanical resonator, which
    in turn is coupled to a resonant electrical LC-circuit.  The
    system can be pumped or read out by both microwave and optical
    drives.}
  \label{EOMcircuit}
\end{figure}

Recent experiments have demonstrated the ability to control mesoscopic
mechanical resonators near the quantum
limit~\cite{Teufel2011,Chan2011,Verhagen2012,Oconnell2010}.  This
achievement provides novel opportunities for fundamental
physics~\cite{Kleckner2008} and a technology for engineering quantum
systems~\cite{Stannigel2012,Hong2012}.  The mechanical resonators are
formally equivalent to electromagnetic resonators, which form basic
elements of quantum optics, but offer many new and unique
opportunities.  Mechanical resonators are massive objects that can be
coaxed into interacting strongly with many different systems.  For
instance, in experiments to date, mesoscopic mechanical objects have
been coupled to electrical and optical photons in cavities, although
not to both simultaneously.  As proposed in Ref.~\cite{Regal}, it may
be possible in the near future to couple a mechanical resonator to
both electrical and optical cavities at the same time
(Fig.~\ref{EOMcircuit}).  Such an interface would provide a way to
connect quantum resources that are more suited for creating and
manipulating quantum states ({\em i.e.}~electrical
circuits)~\cite{Hofheinz2009} to resources that are more suited for
transmitting quantum states ({\em i.e.}~optical platforms).

Given this emerging possibility, an important question is how to
harness mechanical resonators within electromagnetic cavities to
transduce and store quantum states. By transduction, we mean the
transfer of energy between distinct degrees of freedom; in this case,
between electromagnetic oscillators whose frequencies are separated by
many orders of magnitude.  The ability to strongly couple single
photons in optomechanical systems would open up a wide variety of
quantum protocols~\cite{Rabl:PhotonBlockadeOptomechanics}. However, in
order to achieve sufficient coupling, experiments mainly operate the
electromagnetic-mechanical interface in an analogous way to three-wave
mixing in nonlinear optics~\cite{Zhang2003,Akram2010}.  A strong
pump-tone red-detuned from the cavity is introduced to bridge most of
the energy gap between the electromagnetic and mechanical
oscillators. This produces an effective beam-splitter interaction that
can conveniently be turned on and off by varying the pump-tone
intensity.  Single-photon states detuned from the pump can then be
transduced between mechanical excitations and any number of
electromagnetic modes.

The quantum optomechanics experiments envisioned here are thus rooted
in the well developed toolbox associated with two-mode quantum optics.
However, when we introduce low-frequency mechanical resonators, the
presence of a thermal bath damping and exciting the phonon resonances
must be accounted for in the theoretical analysis.  To create a
versatile interface between microwave and optical photons, we consider
use of a megahertz membrane microresonator~\cite{Teufel2011}.  Despite
recent progress toward bringing such mechanical systems to the quantum
regime, the mechanical decoherence rate, which is proportional to the
product of the mechanical resonator line strength and occupation
number of thermal bath phonons, remains a significant decay pathway.

In this paper, we consider the effect of decoherence on the
application of mechanical resonators to store quantum mechanical
states and to transduce states between electromagnetic resonators.
Our analysis is based on a quantum state diffusion (QSD)
method~\cite{Wiseman:QuantumJumpsAndDiffusion} for solving the
evolution of open quantum systems. This method provides an exact
unraveling of the quantum master equation into parallel pure-state
quantum trajectories.  Using the QSD method we are able to calculate
the fidelity of this system for quantum state memory applications and
quantum state transduction.

It should be emphasized that this numerical method is not restricted to
solving the dynamical evolution of only Gaussian or classical states.
The QSD method allows for any quantum input state to be tested.  In
this paper, we compare the memory and transduction fidelity for
coherent states, squeezed-states, Schr\"odinger-cat states, and
nonclassical superpositions of Fock states.  Such numerical approaches
also allow for the analysis of many types of coupling schemes.  The
simplest scheme is the coherent swapping of the quantum state of two
oscillators at a transfer rate determined by the strength of the
coupling.  This is the optomechanics analogue of Rabi flopping between
internal states of a two-level atom.  In addition to this
scheme~\cite{Tian2010}, we explore several more diverse swapping
schemes and find them to be more robust against the omnipresent
mechanical decoherence~\cite{Wang2011,Clerk2012,Tian2012}.

\begin{figure} \includegraphics[width=60mm]{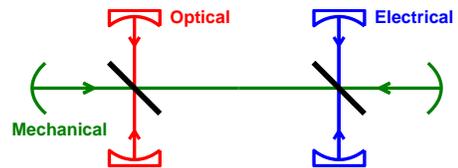}
% \begin{figure} \includegraphics[width=60mm]{pics/INTRO_BeamSplitter.eps}
  \caption{(Color online) An equivalent system of two coupled
    adjustable beam splitters which can be a formal analogue to the
    two-mode optomechanical system in the linearized approximation.
    In this open quantum system, each oscillator is also coupled to
    its respective reservoir (not shown).}
  \label{BeamSplitterDiagram}
\end{figure}

As we will show, the basic system we consider formally corresponds
precisely to a set of adjustable beam-splitters and cavities as
illustrated in Fig.~\ref{BeamSplitterDiagram}.  The optomechanical and
electromechanical coupling strength is represented by the reflectivity
of the effective beam splitters and can be adjusted by variation of
external parameters to be anywhere from 0 to $100 \%$. We first look
at the system from the perspective of utilizing this beam-splitter
interaction to facilitate the storage and retrieval of an
electromagnetic quantum state in a mechanical resonator.  Second, we
look at the system from the perspective of transduction of a quantum
state from a microwave to optical resonator, or vice versa.  We
investigate the effect of different protocols on the population of the
mechanical state and, hence, the susceptibility to mechanical
decoherence.

In all cases, we set the decay of the optical and microwave cavities
equal to zero and consider the state preparation of the optical and
microwave modes as an initial condition.  This simplifies the analysis
and allows us to focus on the role of mechanical decoherence.
Nonetheless, the internal and external $Q$ of the optical and
microwave cavities is also an important topic, and has recently been
treated in both the context of swapping \cite{Tian2010} and itinerant
photon schemes \cite{Wang2011,Safavi-Naeini,Clerk2012}.

%%%%%%%%%%%%%%%%%%%%%%%%%%%%%%%%%%%%%%%%%%%%%%%%%%%%%%%%%%%%%%%%%%%%%%%%%%%%%%%%
\section{Theory}

The coupled electro-opto-mechanical system~\cite{Zhang2003,Akram2010}
is described by the Hamiltonian in the Schr\"{o}dinger picture, $
\hat{H} = \hat{H}_{\text{self}} + \hat{H}_{\text{coupling}} +
\hat{H}_{\text{pump}}$, where:
\begin{eqnarray}
  \hat{H}_{\text{self}} &=& \hbar\omega_{o,c}\ \opad\opa 
  + \hbar\omega_{\mu,c}\ \opbd\opb 
  + \hbar \omega_m \opdd\opd\;, \nonumber\\
  \hat{H}_{\text{coupling}} &=& 
  -\frac{\hbar X_{\text{ZP}}}{2}  \left(\opd  +\opdd  \right)
  \left(g_o\opad\opa + g_{\mu}\opbd\opb\right) \;,\nonumber\\
  \hat{H}_{\text{pump}} &=& \hbar \left(\opa A_o^* e^{i \omega_{o} t} 
    + \opad A_o e^{-i \omega_{o} t} \right) \nonumber\\
  & & {} + \hbar \left(\opb A_{\mu}^* e^{i \omega_{\mu} t}
    + \opbd A_{\mu} e^{-i \omega_{\mu} t} \right) \;.
  \label{eq:fullHschrodinger}
\end{eqnarray}
Here the operators $\opa$, $\opb$, and $\opd$ are the annihilation
operators for a photon in the optical cavity, a photon in the
microwave cavity, and a phonon in the mechanical resonator,
respectively.  The system is driven by classical pump fields with
frequencies $\omega_{(o,\mu)}$ and the bare ({\em i.e.}~uncoupled)
cavity resonance frequencies are $\omega_{(o,\mu),c}$, where~$o$
stands for optical and $\mu$ stands for microwave.  For the mechanical
resonator, the resonance frequency is $\omega_m$ and the harmonic
oscillator length is $X_{\text{ZP}}$. The coupling constants $g_o$ and
$g_\mu$ are physically determined by the amount of the shift of the
resonant frequency of each cavity with respect to changes in the
mechanical resonator position.

The strong coherent pump amplitudes, $A_o$ and $A_\mu$, lead to a
buildup of large steady-state fields in the optical and microwave
resonators, whose purpose is to increase the optomechanical coupling
(a fact which we demonstrate below).  The resulting steady-state
intracavity field amplitudes in turn shift the equilibrium position of
the mechanical resonator through the radiation pressure force. We
begin by finding these steady-state semiclassical solutions. We remove
the time-dependence of the Hamiltonian in
Eq.~(\ref{eq:fullHschrodinger}) by transforming to an interaction
picture rotating at the drive frequency, {\em i.e.}~with $e^{i
  \hat{H_0} t}$ where
\begin{equation}
  \hat{H_0} = \hbar\omega_o\ \opad\opa + \hbar\omega_\mu\ \opbd\opb\,.
\end{equation}
We define detunings as the difference of the drive frequencies from
their respective bare-cavity resonances
\begin{equation}
\Delta_{(o,\mu)}= \omega_{(o,\mu),c} - \omega_{(o,\mu)} \;.
\end{equation}
The interaction Hamiltonian then becomes,
\begin{eqnarray}
  \hat{V_0} &=& \hbar\Delta_o \opad \opa + \hbar \Delta_\mu \opbd \opb
  +  \hbar\omega_m \opdd \opd \nonumber \\
  & & {}
  -\frac{\hbar X_{\text{ZP}}}{2}  \left(\opd  +\opdd  \right)
  \left(g_o\opad\opa + g_{\mu}\opbd\opb\right) \nonumber\\
  & & {}
  +\hbar \left(\opa A_o^* + \opad A_o  \right) 
  + \hbar \left(\opb A_{\mu}^* + \opbd A_{\mu} \right)\,.
\label{fullH}
\end{eqnarray}
From this we derive coupled equations of motion for the macroscopic
fields, $\alpha=\bigl<\opa\bigr>$, $\beta=\bigl<\opb\bigr>$, and
$\delta=\bigl<\opd\bigr>$. These equations are found by writing the
Heisenberg operator equations and directly substituting for each
operator its corresponding classical amplitude:
\begin{eqnarray}
  i\frac{d\alpha}{dt}&=&\left(\Delta_o-i\frac{\kappa_o}2\right)
  \alpha-\frac{g_o X_{\text{ZP}}}2
  \left(\delta+\delta^*\right)\alpha+A_o\,, \nonumber\\
  i\frac{d\beta}{dt}&=&\left(\Delta_{\mu}-i\frac{\kappa_{\mu}}2\right)
  \beta-\frac{g_{\mu} X_{\text{ZP}}}2
  \left(\delta+\delta^*\right)\beta+A_{\mu}\,, \nonumber\\
  i\frac{d\delta}{dt}&=& \omega_m\delta-\frac{X_{\text{ZP}}}2
  \left(g_o|\alpha|^2+g_{\mu}|\beta|^2\right)\,.
\end{eqnarray}
where $\kappa_{o,\mu}$ is the damping of each electromagnetic
oscillator. We introduce damping to insure the system relaxes to
steady-state. We find the steady-state solution by setting the time
derivatives on the left-hand side of each equation to zero. Solving
the last equation shows that the equilibrium position of the
mechanical oscillator is shifted due the force exerted on it by the
radiation pressure in the cavities, with steady-state value
$\delta_{\text{s}}=
X_{\text{ZP}}\left(g_o|\alpha|^2+g_{\mu}|\beta|^2\right)/(2\omega_m)$. Note
this is purely real indicating no shift in the equilibrium momentum of
the mechanical oscillator.

Substituting the steady-state result for $\delta$ into the remaining
two equations for $\alpha$ and $\beta$ produces two coupled algebraic
equations that contain cubic terms in the field amplitudes. These
equations have multiple roots for small $\kappa_{o,\mu}$,
corresponding to the well-known steady-state solutions for two-mode
optical bistability. For both the optical and microwave cavities, we
are interested in the stable solutions on the high-intensity branch,
which we denote by $\alpha_{\text{s}}$ and $\beta_{\text{s}}$
respectively.

Having determined the semiclassical solutions, we now proceed to
consider the fluctuations about these solutions. We linearize all
three fields:
\begin{eqnarray}
  \opa&\rightarrow\alpha_{\text{s}}+\opa\nonumber\\
  \opb&\rightarrow\beta_{\text{s}}+\opb\nonumber\\
  \opd&\rightarrow\delta_{\text{s}}+\opd
\end{eqnarray}
and substitute these into Eq.~(\ref{fullH}). We may then identify the
mean-field energy shift that contains no operators
\begin{eqnarray}
  E &=& \hbar\Delta_o|\alpha_{\text{s}}|^2 + \hbar \Delta_\mu |\beta_{\text{s}}|^2
  +  \hbar\omega_m |\delta_{\text{s}}|^2 \nonumber \\
  & & {}
  -\hbar X_{\text{ZP}}  \delta_{\text{s}}
  \left(g_o|\alpha_{\text{s}}|^2 + g_{\mu}|\beta_{\text{s}}|^2\right) \nonumber\\
  & & {}
  +\hbar \left(\alpha_{\text{s}} A_o^* + \alpha_{\text{s}}^* A_o  \right) 
  + \hbar \left(\beta_{\text{s}} A_{\mu}^* + \beta_{\text{s}}^* A_{\mu} \right)
  \label{eq:offset}
\end{eqnarray}
Using this, the final step in the derivation is to subtract the energy
offset $E$ in Eq.~(\ref{eq:offset}) from the Hamiltonian in
Eq.~(\ref{fullH}) and transform the resulting interaction into an
appropriate rotating frame. Since at this point we are working in the
interaction picture rotating at the drive frequency, we now need to
make a second interaction picture transformation, that is, on top of
the previous one, to transform into a rotating frame at the cavity
frequencies for all oscillator modes; optical, microwave, and
mechanical. In doing this we must take careful account of the
radiation pressure shift that we have just derived for the
electromagnetic resonant frequencies. We thus perform a second
transformation into an interaction picture rotating with $e^{i
  \hat{H_1} t}$ where
\begin{equation}
  \hat{H_1} \equiv \hbar\tilde{\Delta}_o\ \opad\opa 
+ \hbar\tilde{\Delta}_\mu\ \opbd\opb
  + \hbar\omega_m\opdd\opd \,.
\end{equation}
and we have defined $\tilde{\Delta}_o=\Delta_o-X_{\text{ZP}}\delta
g_o$ and $\tilde{\Delta}_{\mu} =\Delta_{\mu}-X_{\text{ZP}}\delta
g_{\mu}$ in order to account for the aforementioned shift in the
cavity frequency due to radiation pressure. This leads to the
interaction Hamiltonian
\begin{eqnarray}
  \hat{V_1} &=& -\frac{\hbar X_{\text{ZP}}g_0}2\Bigl(
  \opd\opad\alpha_{\text{s}}e^{i(\tilde{\Delta}_o-\omega_m)t}
  + \opd\opa\alpha_{\text{s}}^*e^{-i(\tilde{\Delta}_o+\omega_m)t} \nonumber\\
  &&{}\quad
  + \opdd\opad\alpha_{\text{s}}e^{i(\tilde{\Delta}_o+\omega_m)t}
  + \opdd\opa\alpha_{\text{s}}^*e^{-i(\tilde{\Delta}_o-\omega_m)t}\Bigr) \nonumber\\
  &&{}-\frac{\hbar X_{\text{ZP}}g_{\mu}}2\Bigl(
  \opd\opbd\beta_{\text{s}}e^{i(\tilde{\Delta}_{\mu}-\omega_m)t}
  + \opd\opb\beta_{\text{s}}^*e^{-i(\tilde{\Delta}_{\mu}+\omega_m)t} \nonumber\\
  &&{}\quad
  + \opdd\opbd\beta_{\text{s}}e^{i(\tilde{\Delta}_{\mu}+\omega_m)t}
  + \opdd\opb\beta_{\text{s}}^*e^{-i(\tilde{\Delta}_{\mu}-\omega_m)t}\Bigr)
\end{eqnarray}
From this result, we can see that in order to maximize the
beam-splitter couplings, the strong pump fields should be detuned to
the red of their respective microwave or optical cavity by
\begin{equation}
  \tilde{\Delta}_o=\tilde{\Delta}_\mu=\omega_m\;.
\end{equation}
In the resolved sideband limit where the frequency, $\omega_m$, of the
mechanical resonator is much larger than the mechanical decay rate as
well as the effective coupling constants, we can then employ the
rotating wave approximation where all the rapidly oscillating time
dependent terms that contain $e^{2i\omega_m}$ average to zero. Putting
all this together leads to the following effective Hamiltonian
\begin{align} 
  \hat{H}_{\text{eff}} =& \hbar\omega_m \left(\opad \opa + \opbd \opb
    + \opdd \opd\right)\nonumber \\
  & -\hbar\frac{\Omega_o}{2} \left(\opad \opd + \opa\ \opdd\right)
  -\hbar\frac{\Omega_{\mu}}{2} \left(\opbd\opd + \opb\ \opdd\right)\;.
  \label{eq:beamSplitterH}
\end{align}
where the modified coupling constants are now 
\begin{eqnarray}
\Omega_{o}&=&g_{o}
X_{\rm ZP} \alpha_{\text{s}}\,,\nonumber\\
\Omega_{\mu}&=&g_{\mu} X_{\rm ZP}
\beta_{\text{s}}\,.
\end{eqnarray}
Without loss of generality, we have taken both $\alpha_{\text{s}}$ and
$\beta_{\text{s}}$ to be real. The largest values achieved in
experiment are about $\Omega_{o}\sim 0.1 \omega_m$ and
$\Omega_{\mu}\sim 0.1 \omega_m$.~\cite{Teufel2011:couplingNumbers,
  Verhagen2012}

As noted earlier, this bilinear Hamiltonian is analogous to three
quantized single mode fields coupled to each other by beam splitters.
The beam splitters are adjustable by adjusting the $\Omega$'s.  A
coupling of $\pi/2$ will be like a 50/50 beam splitter.  A coupling of
$\pi$ will be like a mirror, swapping the states perfectly.  If we
turn the coupling off, the oscillators will propagate freely as if
there is no beam splitter.  Thus, by varying the coupling constants,
we can change from 0 to $100 \%$ reflection and transmission.  It is
important to note that the quantum mechanical systems described by the
field operators $\hat a$, $\hat b$, and $\hat d$ are really
fluctuations of the bare fields around their stationary values at each
of their respective resonator frequencies.

%%%%%%%%%%%%%%%%%%%%%%%%%%%%%%%%%%%%%%%%%%%%%%%%%%%%%%%%%%%%%%%%%%%%%%%%%%%%%%%%
% \section{Numerical Methods}

In addition to the Hamiltonian in Eq.~\ref{eq:beamSplitterH}, the resonators are subject to dissipative processes that stem from their coupling to the environment.  The mechanical resonator is coupled to a thermal bath.  Thermal phonons are absorbed into the mechanical resonator at a rate $\gamma_m \bar{n}$ where $\bar{n}$ is the average Bose occupancy of the resonator and phonons are lost at a rate $\gamma_m (\bar{n}+1)$.  The optical and microwave cavities are taken to be at zero temperature, and we set the loss rates given by $\gamma_o$ and $\gamma_\mu$ respectively to zero.

We use the quantum state diffusion (QSD) method to simulate the fully quantum evolution of this open  system~\cite{Gisin:QuantumStateDiffusion, Persival:QuantumStateDiffusion, Belavkin:QuantumStateDiffusion, Wiseman:QuantumJumpsAndDiffusion}.  The QSD method is well suited for this problem because it yields the conditional evolution of an open quantum system subject to homodyne measurements of the output fields.  In this way, we may obtain amplitude and phase information about the decay channel.  In the QSD approach, we stochastically evolve each resonator subsystem as if we were performing a continuous fictitious homodyne measurement of the photons or phonons coming out of each resonator.  Even though an actual experiment can not measure the phonons in the mechanical resonator, the numerical simulation gives us access to this information in the spirit of a `Gedanken' measurement.  

For the scenarios we explore, the decay rates of the optical and microwave resonators are set to zero, so that only the mechanical resonator is coupled to the environment.  Thus, in those cases, the stochastic evolution only applies to the mechanical resonator subsystem, while the optical and microwave resonators evolve according to the Schr\"{o}dinger equation.  In actual experiments, the optical and microwave resonators are also coupled to the environment and actual homodyne measurements can be performed on those output fields.  The QSD method works for the fully open system as well.  But here, we are reducing the fully open system to focus only on the mechanical decoherence.

In the QSD method, the evolution of the total density matrix of the system is unraveled into an ensemble of stochastic parallel pure state trajectories.  Each trajectory is evolved according to a stochastic differential equation.  The trajectories are then averaged in the ensemble sense to recreate the total density matrix.  In the limit of a large number of trajectories, the ensemble average of the stochastic trajectories goes to a state diffusion evolution.  Each trajectory evolves according to the stochastic differential equation~\cite{Wiseman:QuantumJumpsAndDiffusion}
\begin{align}
  \ket{\tilde{\Psi}(t+dt)} =& \Bigg\{1 -\bigg[\frac{i}{\hbar}\hat{H}_{\text{eff}} 
     + \frac{\gamma_m (2\bar{n}+1)}{2}\opdd\opd \nonumber\\
     & \quad\qquad + 2 \gamma_m (2\bar{n}+1) \langle \opdd + \opd \rangle \opd \bigg]dt\nonumber\\
     &\quad + \opd \sqrt{\gamma_m \bar{n}}\ dW_{\opd}(t) \nonumber\\
     &\quad + \opdd \sqrt{\gamma_m (\bar{n}+1)}\ dW_{\opdd}(t) \Bigg\} \ket{\Psi(t)}
\end{align}
where, $dW(t)$ is the continuum limit of a Wiener increment, $\Delta W$, which satisfies the ensemble average $\langle (\Delta W)^2\rangle = \Delta t$ of a Gaussian random distribution with a width $\Delta t$.  There are two Wiener increments, one for each noise process in the mechanical oscillator where there are two types of decay channels, one for phonons entering the system, $dW_{\opdd}(t)$, and one for phonons leaving the system, $dW_{\opd}(t)$.  We numerically integrate these stochastic differential equations using a second order scheme~\cite{KloedenAndPlaten}.

%%%%%%%%%%%%%%%%%%%%%%%%%%%%%%%%%%%%%%%%%%%%%%%%%%%%%%%%%%%%%%%%%%%%%%%%%%%%%%%%
%\section{Results}

%%%%%%%%%%%%%%%%%%%%%%%%%%%%%%%%%%%%%%%%%%%%%%%%%%%%%%%%%%%%%%%%%%%%%%%%%%%%%%%%
\section{Quantum State Memory}

One of the possible applications of this system is to store a quantum state in the mechanical resonator.  One could prepare the microwave resonator in any of a variety of quantum states. This can be done with superconducting quantum circuits or other experimental setups such as those described in~\cite{genOptQuState, genEMQuState, squid1, Hofheinz2009}.  Then, the states of the microwave and mechanical resonators could be swapped using a ``$\pi$-pulse'', effectively storing the quantum state in the mechanical resonator.  At some later time, another swap could be done to put the state back into the microwave resonator where it can be retrieved~\cite{Zhang2003, Tian2010}.  The objective would be to maintain high fidelity of the quantum state involved.  These swaps are achieved by varying the coupling constant, $\Omega_{(o,\mu)}$ which behaves like a Rabi frequency in the beam splitter Hamiltonian in Eq.~\ref{eq:beamSplitterH}.  The coupling constants can be changed by modulating the bare coupling constants $g_{(o,\mu)}$ or the complex pump amplitude $A_{(o,\mu)}$, or the detuning $\Delta_{(o,\mu)}$.  For this problem, it is sufficient to consider just a pair of resonators.  We reduce the system to one electromagnetic resonator and one mechanical resonator by setting one of the coupling constants to zero.

\begin{figure}
  \includegraphics[width=85mm]{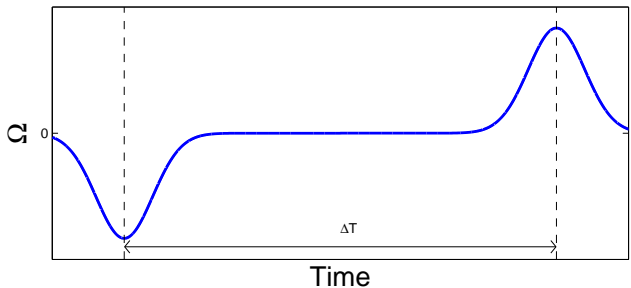}
  \caption{A schematic diagram of the coupling $\pi$ pulse sequence for Quantum State Memory tests.  The sign of the coupling constant for retrieval must be the opposite of the sign for storage in order to cancel the phase accumulation during the pulses.}
  \label{MemoryPulseSeq}
\end{figure}

A ``$\pi$-pulse'' in this context is a pulse where the time integral over the coupling constant in frequency units is $\pi$.  For the Gaussian coupling pulses, $\Omega(t) = \Omega e^{-(t-t_c)^2/(w^2 \pi)}$, that we employ, the pulse area is $w \pi \Omega$ where $w\pi$ is the width of the Gaussian and the peak amplitude is $\Omega$.  The pulse sequence is schematically shown in Fig.~\ref{MemoryPulseSeq}.  

To see how well the system is able to store classical and quantum states, we study a variety of initial states with the same pulse sequence for storage and retrieval.  Recall that these states in fact represent the phonon or photon fluctuations around the stationary value of each resonator, as we have previously described in the derivation of the linearized effective Hamiltonian in Eq.~\ref{eq:beamSplitterH}.

We use coherent states $\ket{\alpha}\ket{n}_m$, squeezed coherent states $\ket{\alpha,\xi}\ket{n}_m$, cat states $\ket{\psi_{cat}}$, and a superposition of Fock states $\ket{\psi_{SF}}$ as inputs to the microwave resonator.  The squeezed coherent state~\cite{Walls:squeezed} is a displaced squeezed vacuum state, $\ket{\alpha,\xi}=\hat{D}(\alpha)\hat{S}(\xi)\ket{0}$, where $\alpha$ is the mean value and $\xi$ is the squeezing parameter. The displacement operator is $\hat{D}(\alpha)=e^{\alpha \opad-\alpha^* \opa}$ and the squeezing operator is $\hat{S}(\xi)=e^{\frac{1}{2}(\xi^* \opa^2-\xi (\opad)^2)}$.

The cat state is $\ket{\psi_{cat}} = N(\ket{\alpha}+\ket{-\alpha})\ket{n}_m$ where $N$ is a normalization constant.  The input state for the superposition of Fock states that we will study is $\ket{\psi_{SF}} = \frac{1}{\sqrt{2}} (\ket{0}_\mu + \ket{1}_\mu) \ket{n}_m$, where $\ket{i}_\mu$ indicates the photon fluctuations around the stationary value inside the microwave cavity and $\ket{n}_m$ indicates the phonon fluctuations around the stationary value inside the mechanical resonator.  

The initial Fock state for the mechanical resonator for all input states, $\ket{n}_m$, is randomly chosen according to the probability distribution for a thermal state, 
\begin{equation}
  P_n=\frac{\bar{n}^n}{(\bar{n}+1)^{n+1}}\;,
  \label{eq:thermalProb}
\end{equation}
in order to sample the thermal density matrix at a temperature corresponding to $\bar{n}$.

To measure the success of this scheme, we look at the fidelity~\cite{Tian2010} of retrieving the input state as a function of the mechanical quality factor, $Q_m=\omega_m/\gamma_m$.  The fidelity~\cite{fidelity} is defined as 
\begin{equation}
  F(\rho_i,\rho_f) = \left[Tr\left(\sqrt{\sqrt{\rho_i}\rho_f\sqrt{\rho_i}}\right)\right]^2
  \label{eq:origFidelity}
\end{equation}
where $\rho_i$ and $\rho_f$ are the reduced density matrices for the input and output states respectively.  The fidelity for pure states reduces to the overlap between the initial and final states. 

At zero temperature, in the case where the pulse duration is short compared to the decay time of the mechanical resonator, we can neglect the decay during the swapping pulses and assume that the swaps happened perfectly.  In that case, the final state will only have decayed exponentially, at the mechanical decay rate $\gamma_m$, during the time, $\Delta T$, between the two $\pi$ pulses to become $\ket{\psi_f} = \ket{e^{-\gamma_m  \Delta T/2} \alpha}$.  Thus we can analytically find the fidelity for a pure final state to be
\begin{equation}
  F = |\langle e^{-\frac{\omega_m \Delta T}{2 Q_m}} \alpha |  \alpha\rangle |^2 
    = e^{-| \alpha|^2(1-e^{-\omega_m \Delta T/(2 Q_m)})^2}\;.
  \label{eq:CoherentStateFidelity0}
\end{equation}

% At zero temperature, in the case where the pulse duration is short compared to the decay times of the mechanical resonator, we can analytically find the fidelity for a pure final state.  For the coherent case, the initial state is $\ket{\alpha}$ and the final state has simply decayed exponentially during the time, $\Delta T$, between the two $\pi$ pulses to be $\ket{\psi_f} = \ket{e^{-\gamma_m  \Delta T/2} \alpha}$.  Thus, we find the fidelity to be
% \begin{equation}
%   F = |\langle \alpha | e^{-\frac{\omega_m \Delta T}{2 Q_m}} \alpha \rangle |^2 
%     = e^{-|\alpha|^2(1-e^{-\omega_m \Delta T/(2 Q_m)})^2}\;.
%   \label{eq:CoherentStateFidelity0}
% \end{equation}

However, for non-zero temperatures or large decay rates the final state will thermalize quickly and this formula will no longer be valid.  The state does not decay to the vacuum as Eq.~\ref{eq:CoherentStateFidelity0} suggests.  Rather it decays to a thermalized value.  Thus, the final fidelity for pure states will saturate at low-Q to the overlap between a thermalized state and the initial state.  Thus, the coherent state fidelity takes on the form,
\begin{equation}
  F(Q) = F(0) + \left(1-F(0)\right) e^{-|\alpha|^2(1-e^{-\omega_m \Delta T/(2 Q_m)})^2}\;.
  \label{eq:CoherentStateFidelity}
\end{equation}
We take the overlap between the thermal state given by Eq.~\ref{eq:thermalProb} and the coherent state in the number basis to obtain the thermal saturation overlap value,
\begin{equation}
  F(0) = \frac{e^{-|\alpha|^2/(1+\bar{n})}}{1+\bar{n}} \;.
\end{equation}

Further complicating matters, the full density matrix, which is pure, must be reduced to the resonator subsystem that is being read out before taking the fidelity overlap with the reduced input density matrix.  This makes the reduced density matrices impure.  Consequently, we need to employ the more general formula to find the fidelity that is valid for states that are not pure.  

\begin{figure}
  \includegraphics[width=85mm]{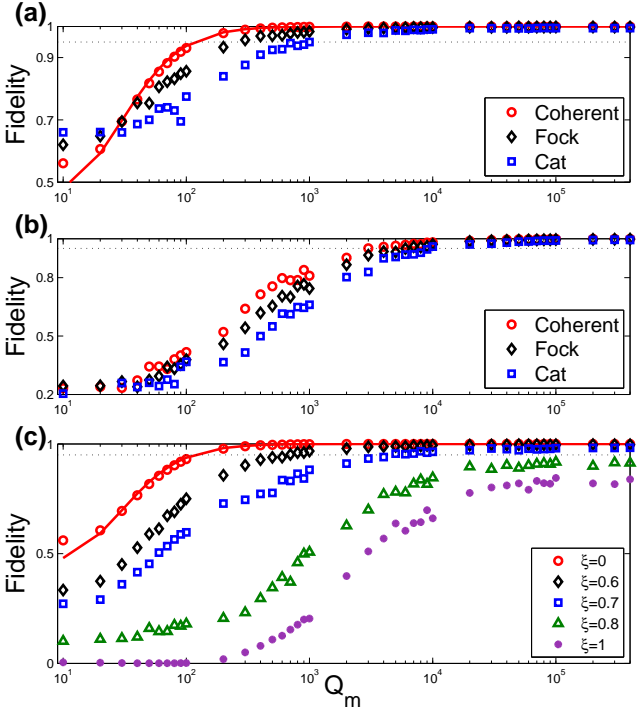}
  \caption{(Color online) Memory fidelity as a function of mechanical resonator quality. (a): For $\bar{n}=0$.  Input states $\ket{\alpha}$ (red circles) with $\alpha=1$, $\ket{\psi_{SF}}$ (black diamonds),
  $\ket{\psi_{cat}}$ (blue squares).  The solid line is the analytic formula for the coherent fidelity from Eq.~\ref{eq:CoherentStateFidelity}.  (b): For $\bar{n}=3$. (c): For squeezed states $\ket{\alpha, \xi}$ with $\alpha=1$ and $\bar{n}=0$ for various squeezing parameters $\xi$.  In all cases, ${\Delta}T=64 (units of \frac{1}{\omega_m})$ and the peak coupling is $\Omega_\mu=0.1\omega_m$.  The horizontal dotted black line indicates 95\% fidelity for reference.}
  \label{MemoryAllStates}
\end{figure}

Fig.~\ref{MemoryAllStates}A shows the memory fidelity at zero temperature for the input states $\ket{\alpha}$, $\ket{\psi_{SF}}$, and $\ket{\psi_{cat}}$ for $\Delta T=64(units of 1/\omega_m)$.  The coherent state fidelity agrees well with our analytical result in Eq.~\ref{eq:CoherentStateFidelity} except at very low Q.  At low Q values, the input state thermalizes quickly to a thermal state.  For these low Q values, the decay time is smaller than the pulse width.  Even though there is almost no time between the pulses for these cases, the state still thermalizes during the pulse width.  A thermal state has a constant fidelity overlap with the initial state.  Thus, the fidelity for low Q mechanical resonators levels off to a constant value.  This causes the actual fidelity to deviate from the analytic formula.  Across the range of $Q_m$ values considered in Fig.~\ref{MemoryAllStates}, the non-classical states have lower fidelities than the coherent state.

The fidelities for a finite temperature corresponding to $\bar{n}=3$ are shown in Fig.~\ref{MemoryAllStates}B.  As expected, the fidelities are consistently reduced by the thermal noise.  At the current experimental values for the mechanical quality~\cite{Teufel2011:couplingNumbers} (far right hand side of the plots), all the fidelities are above $99 \%$ and there is little difference between the various input states.

\begin{figure}
  \includegraphics[width=85mm]{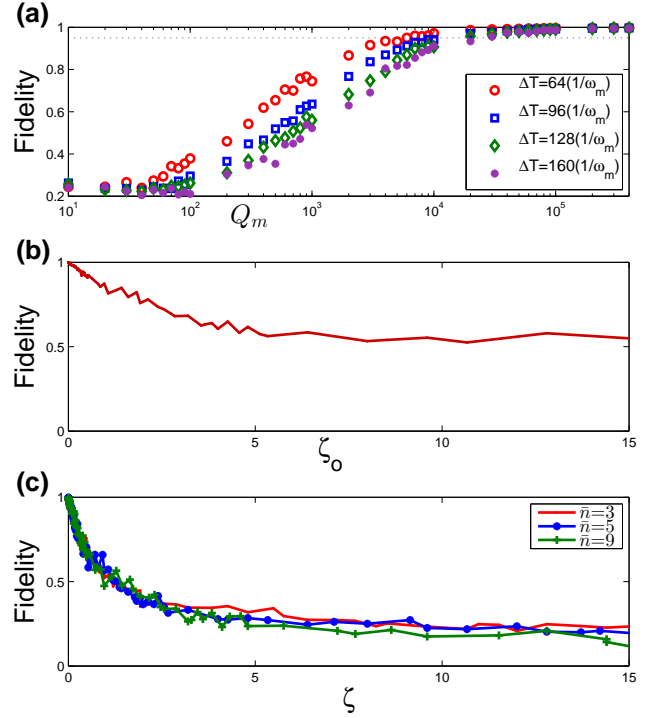}
  \caption{(Color online) (a): Memory fidelity of $\ket{\psi_{SF}}$ as a function of mechanical resonator quality for various wait times for $\bar{n}=3$. The fidelity decreases exponentially with decreasing Q and with increased wait times. At low-Q, the fidelity saturates to the thermalized value. (b): The memory fidelity vs. scaled wait time for $\bar{n}=0$. (c): The memory fidelity vs. scaled wait time for different temperatures showing how all the curves collapse onto one universal curve. In all cases, the peak coupling is $\Omega_\mu=0.1\omega_m$.}
  \label{MemoryWaitPulses}
\end{figure}

For the squeezed states $\ket{\alpha,\xi}$, shown in Fig.~\ref{MemoryAllStates}C, we varied the squeezing parameter $\xi$, keeping $\alpha$ constant at $\alpha=1$. For $\xi=0$ the input state is just the coherent state, so the data is similar to the coherent state data.  The solid red line is the analytic formula from Eq.~\ref{eq:CoherentStateFidelity} showing good agreement with the no squeezing, $\xi=0$, input state.  As the squeezing increases, the input state becomes more and more nonclassical.  For low-Q cavities, the mechanical resonator quickly decays to a thermal state that is farther and farther away from the initial squeezed state.  This causes the fidelity for these highly squeezed states to go to zero.

Next, we look at how long the mechanical oscillator can store a quantum state before significant degradation occurs.  Fig.~\ref{MemoryWaitPulses}A shows the memory fidelity of $\ket{\psi_{SF}}$ for increasing wait times as a function of $Q_m$ for $\bar{n}=0$.  As expected, the fidelity decreases with increasing wait time, but we can still achieve above a 95\% fidelity for the higher $Q_m$ values.  At the current experimental Q values of about $Q_m=360,000$~\cite{Teufel2011:couplingNumbers} the quantum state can be stored in the mechanical resonator for longer than ${\Delta}T=160 (units of 1/\omega_m)$ at low temperatures.

For the coherent case, we have a universal curve for the zero temperature fidelity, 
\begin{equation}
  F = e^{-|\beta(0)|^2(1-e^{-\zeta_o/2})^2} \;.
\end{equation}
where, we have rescaled the fidelity data by a dimensionless variable scaled by the thermal average occupation number $\bar{n}$, 
\begin{align}
  \zeta_o =& \omega_m (\Delta T-\frac{\pi}{\Omega_\mu})/Q_m \\
  \zeta =& \omega_m (\Delta T-\frac{\pi}{\Omega_\mu}) \bar{n}/Q_m \;,
\end{align}
For the Fock states, the exponential dependence on $\zeta$ is similar, although we no longer have an analytic formula.  We subtract the width of the $\pi$ pulses to more accurately reflect the actual storage time.  Fig.~\ref{MemoryWaitPulses}B shows the same memory fidelity as for subplot~A versus $\zeta_o$ for $\bar{n}=0$.  Fig.~\ref{MemoryWaitPulses}C shows the memory fidelity versus $\zeta$ for finite $\bar{n}$ in order to remove the dependence on the temperature.  As we increase the wait time before the second $\pi$ pulse, the fidelity decreases exponentially as expected.  By removing the dependence on the temperature, the fidelity curves all collapse onto one universal curve.  Even though we are running our simulations at low $\bar{n}$ values, these results can be scaled up to more experimentally practical $\bar{n}$ values while keeping $\zeta$ constant.  

All the input states in Figure~\ref{MemoryAllStates} reach above a 95\% fidelity for $\zeta_o$ values below about 0.26 for coherent states, 0.11 for Fock states, and 0.04 for Cat states for $\bar{n}=0$.  The fidelity is above 95\% for $\zeta$ values below 0.03 for coherent states, 0.02 for Fock states, and 0.01 for Cat states for $\bar{n}=3$ with the peak coupling set to $\Omega_\mu=0.1\omega_m$.  

\begin{figure}
  \includegraphics[width=85mm]{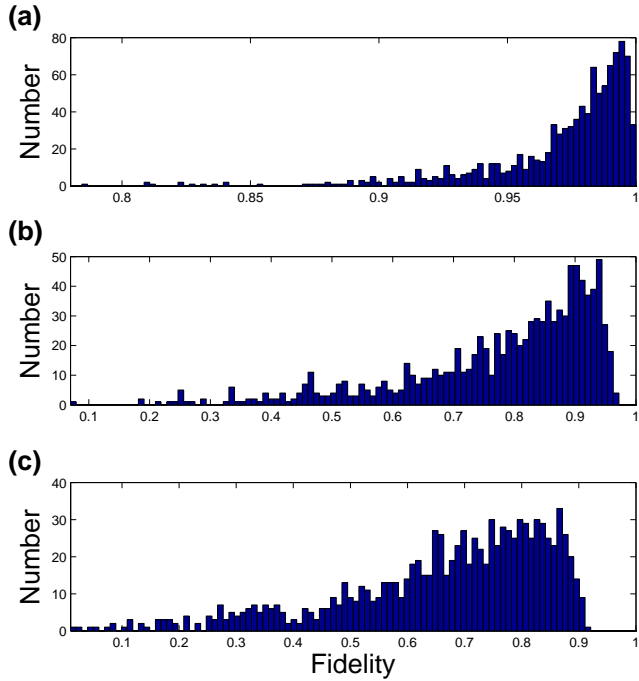}
  \caption{ (Color online) Distribution of fidelities for $\ket{\psi_{SF}}$
  for $Q_m=10,000$ (a): $Q_m=1000$, (b): and $Q_m=500$, (c): for 1000 independent 
  QSD trajectories.  In all cases, ${\Delta}T=64(units of \frac{1}{\omega_m})$, $\bar{n}=3$,
  and the peak coupling is $\Omega_\mu=0.1\omega_m$.  Note the different horizontal axis in each plot.}
  \label{MemoryDistribution}
\end{figure}

The distribution of the fidelities over the individual stochastic QSD trajectories is not Gaussian and thus the mean and variance are not representative of the outcome of a single run.  In an experiment, many runs must be performed to get good statistics about the memory transfer fidelity.  However, as the mean fidelity increases, the distribution becomes more and more narrow and thus more repeatable.  Fig.~\ref{MemoryDistribution} shows the fidelity distribution for $\ket{\psi_{SF}}$ for $Q_m=10,000$, $Q_m=1000$, $Q_m=500$,  where the wait time is ${\Delta}T=64(units of \frac{1}{\omega_m})$, $\bar{n}=3$, and the peak coupling is $\Omega_\mu=0.1\omega_m$ for 1000 trajectories.

%%%%%%%%%%%%%%%%%%%%%%%%%%%%%%%%%%%%%%%%%%%%%%%%%%%%%%%%%%%%%%%%%%%%%%%%%%%%%%%%
\section{Quantum State Transduction}

We turn now to the problem of transducing the quantum state from the microwave domain to the optical domain via the mechanical resonator or vice versa.  This situation is formally equivalent to the quantum memory case if the electromechanical coupling pulse and the optomechanical coupling pulse do not overlap.  This is because transduction can be realized by two sequential steps from one resonator to the mechanical resonator and then to the second resonator, playing the same roles as storage and retrieval in the previously studied case of quantum memory.  However, with three resonators, more varied protocols are possible.  Although we specialize our analysis to the case where the initial state is in a superposition of Fock states, which now also contains the optical vacuum state, $\ket{\psi_{t}}=\frac{1}{\sqrt{2}}(\ket{0}_\mu + \ket{1}_\mu)\ket{n}_m\ket{0}_o$, these procedures could be similarly applied to consider other quantum states.

The coupling parameters can be of similar orders of magnitude for optical and microwave cavities. So, to simplify the analysis, we set the respective optical and microwave coupling parameters equal to each other, $\Omega_o=\Omega_\mu=0.1\omega_m$.  In addition, we set both optical and microwave detunings equal to the mechanical resonator frequency.

%%%%%%%%%%%%%%%%%%%%%%%%%%%%%%%%%%%%%%%%%%%%%%%%%%%%%%%%%%%%%%%%%%%%%%%%%%%%%%%%
% \subsubsection{Separated $\pi$ Pulses}

For transduction, we want to minimize the decay and thermal effects caused by leaving the state in the mechanical resonator for any appreciable time period.  The simplest method to accomplish this is to move the quantum state through the mechanical resonator as quickly as possible.  This method would be good for applications such as quantum information processing where speed is desirable. 

Another method is to adiabatically move the state from the microwave to the optical resonator or vice versa without fully populating the mechanical resonator.  Naturally, adiabaticity requires longer times, but it is also less susceptible to variations in the pulse profiles.  We will discuss this method in more detail in the next section.

\begin{figure}
  \includegraphics[width=85mm]{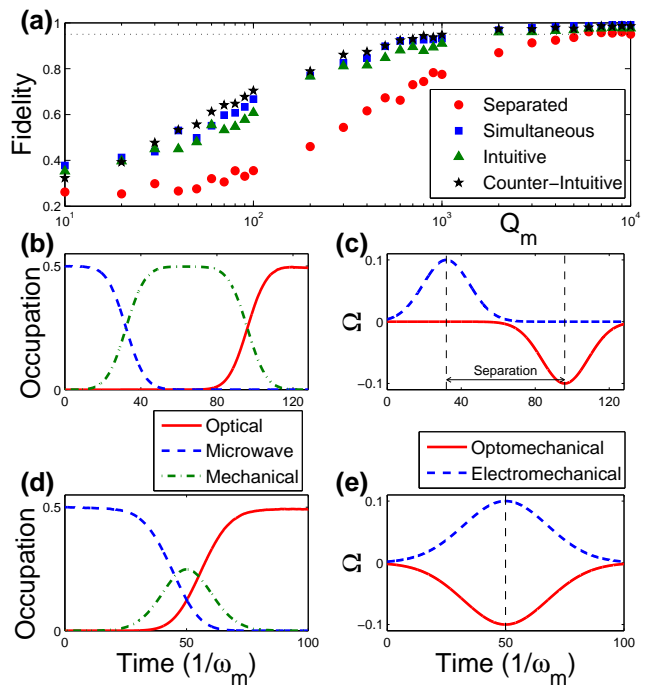}
  \caption{(Color online) (a): Transfer fidelity vs. mechanical resonator quality for $\ket{\psi_{t}}$ for the separated pulse scheme (red dots), the simultaneous pulse scheme (blue squares), intuitive pulse scheme (green diamonds), and counter-intuitive pulse scheme (black stars) with $\bar{n}=3$.  (b): Number of photons or phonons in each resonator for intuitive, separated coupling pulses for zero temperature where ${\Delta}T=64 (units of \frac{1}{\omega_m})$.  (c): The pulse profile for separated pulses.  (d): Number of photons or phonons in each resonator for simultaneous coupling pulses for zero temperature.  (e): The pulse profile for simultaneous pulses.  In all cases, the peak couplings are  $\Omega_o=\Omega_\mu=0.1\omega_m$.}
  \label{TransferPulses}
\end{figure}

We begin by using a similar protocol to the quantum memory scheme.  We set the first $\pi$ pulse to swap the state from the microwave cavity to the mechanical resonator.  Then, we set the second $\pi$ pulse to occur right after the first one to swap the state from the mechanical resonator to the optical cavity.  This pulse sequence is shown schematically in Fig.~\ref{TransferPulses}C.

The resulting fidelity is shown by the red dots in Figure~\ref{TransferPulses}A as a function of the Q of the mechanical resonator.  As this is formally equivalent to the memory scheme, we see similar behavior.  The separation between the peaks of the $\pi$ pulses used here is the same as the wait time we used for the quantum memory scheme.  Thus, the numerical data are similar.  Just like in the quantum memory case,  as the mechanical quality decreases, the thermal noise and decay processes become more significant and the fidelity exponentially decays  down to the fully thermalized value.

%%%%%%%%%%%%%%%%%%%%%%%%%%%%%%%%%%%%%%%%%%%%%%%%%%%%%%%%%%%%%%%%%%%%%%%%%%%%%%%%
% \subsubsection{Simultaneous Pulses}

A natural extension of this scheme is to move the two $\pi$ pulses closer together, which could allow for faster transfer.  Taking this to its logical extreme, we study a scheme where both coupling pulses occur simultaneously.  This allows for the state to move from the microwave to optical cavity, or vice versa, without fully occupying the mechanical resonator, but note, that this in not in the adiabatic regime. 

The overlap in the coupling modifies the optimal pulse area of both couplings. The effective Rabi frequency for the beam splitter Hamiltonian in Eq.~\ref{eq:beamSplitterH} is 
\begin{equation}
\tilde{\Omega} = \sqrt{\Omega_o^2 + \Omega_\mu^2}\;.
\end{equation}
When the optical coupling is turned off, as in the Quantum Memory case of the last section, the effective Rabi frequency reduces to $\tilde{\Omega} = \Omega_\mu$ and the pulse area for each swapping pulse is $\pi$.  However, when both couplings are on and equal in magnitude, the effective Rabi frequency becomes, $\tilde{\Omega} = \sqrt{2} \Omega_o = \sqrt{2} \Omega_\mu$.  Then the pulse area of both pulses increase to $\sqrt{2}\pi$.  This is shown schematically in Fig.~\ref{TransferPulses}E.

This scheme achieves significantly higher fidelities than the separated pulse scheme for all Q values.  The fidelity is shown in Fig.~\ref{TransferPulses}A by the blue squares.  In the low-Q regime, the decay and thermal noise processes become increasingly more significant.  However, the population going through the mechanical resonator is smaller, so the effect is lessened.  Also, there is no waiting time between the pulses for decay and thermal noise processes to occur.  The only decay happens during the width of the pulse.  Thus, the simultaneous pulse scheme is more robust against these processes.

To illustrate the differences between the separated and simultaneous pulse schemes, we examine the populations in each of the three resonators throughout the swapping process.  Fig.~\ref{TransferPulses}B shows how the populations change for the separated $\pi$ pulse scheme.  As expected, the state in the electrical resonator moves into the mechanical resonator and then into the optical resonator.  In contrast, Fig.~\ref{TransferPulses}D shows populations for the simultaneous $\pi$ pulse scheme.  Here, the state transfers from the electrical resonator to the optical resonator without ever fully populating the mechanical resonator.

%%%%%%%%%%%%%%%%%%%%%%%%%%%%%%%%%%%%%%%%%%%%%%%%%%%%%%%%%%%%%%%%%%%%%%%%%%%%%%%%
% \subsubsection{Partially Overlapping Pulses}

As the pulses move closer together, the pulse area needed to make the swap smoothly changes from $\pi$ to $\sqrt{2}\pi$. Also, in real experiments, there may be slight imperfections in the pulse preparation that would result in varying pulse areas and peak separations.  We have run our simulations over a range of varying pulse areas and separations to investigate the effect this had on the fidelity.

\begin{figure}
  \includegraphics[width=85mm]{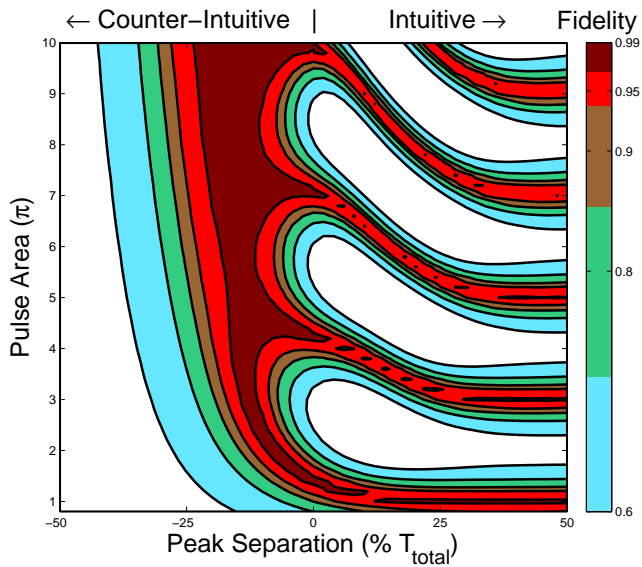}
  \caption{ (Color online) The fidelity vs. pulse area and pulse separation for $Q_m=100,000$, $\bar{n}=0$, and $\Omega_o=\Omega_\mu=0.1\omega_m$.  The negative horizontal axis values represent the peak separation for counter-intuitively ordered pulses while the positive horizontal axis values represent the peak separation for intuitively ordered pulses.}
  \label{EOM_Contours}
\end{figure}

Fig.~\ref{EOM_Contours} shows the transduction fidelity versus the pulse area and the peak separation between the two coupling pulses for $\bar{n}=0$ and $Q_m=100,000$ for the superposition of Fock states, $\ket{\psi_{t}}$.  The horizontal axis represents the separation between the peaks of the two coupling pulses as a percentage of the total transduction sequence time as shown in Figure~\ref{TransferPulses}C.  The zero on the horizontal axis is the point where the two coupling pulses occur simultaneously.  The positive horizontal axis represents peak separations that are ``intuitive'', \textit{i.e.}, the electromechanical coupling pulse occurs before the optomechanical coupling pulse.

As the intuitive Gaussian pulses move farther apart, they eventually become effectively separated and the distance between them no longer matters, since this is at zero temperature with very low decay rates.  This is the area on the far right-hand side of Fig.~\ref{EOM_Contours} where the regions of high fidelity level off at odd integer multiples of $\pi$.  Pulses that are $\pi$ pulses or an odd integer multiple of a $\pi$ pulse will swap the state completely from the microwave resonator to the optical resonator.  Any other pulse area will not perfectly swap the states. Thus, we see the oscillatory behavior we expect in that part of the plot.

At zero pulse separation, the pulses are identical and the Rabi swapping pulse area is $\sqrt{2}\pi$.  Thus, the peak fidelity oscillations are at odd integer multiples of $\sqrt{2}\pi$.  For partially overlapping intuitive pulses, the peak fidelity oscillations smoothly drop from the simultaneous values to the separated values.

The fidelity for a representative partially overlapping intuitive coupling pulse configuration is shown in Fig.~\ref{TransferPulses}A by the green diamonds.  For this case, the pulse area is $1.2 \pi$, the peak separation is $10 \%$, and $\bar{n}=3$.  In the next section, we will discuss the counter-intuitive half of the plot.

%%%%%%%%%%%%%%%%%%%%%%%%%%%%%%%%%%%%%%%%%%%%%%%%%%%%%%%%%%%%%%%%%%%%%%%%%%%%%%%%
\subsection{Adiabatic State Transfer}

The left half of Fig.~\ref{EOM_Contours} shows the transduction fidelity for overlapping coupling pulses that are in the counter-intuitive order, which means, that the optomechanical coupling pulse occurs before the electromechanical coupling pulse.  As the peak separation increases, at some point, the pulses become effectively separated and the fidelity goes to zero.  However, when the coupling pulses are significantly overlapping, there is a large area of high fidelity for any pulse area.  As we increase the pulse area and thus the adiabaticity, the zone of high fidelity transduction increases.  

\begin{figure}
  \includegraphics[width=80mm]{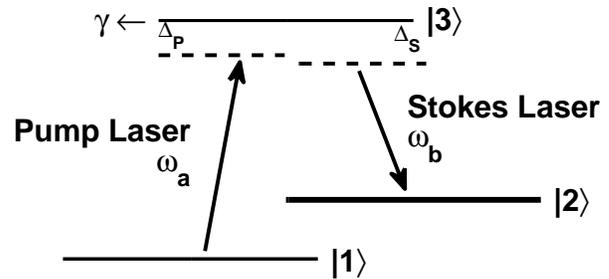}
  \caption{Schematic of a three state system.}
  \label{ThreeLevelAtom}
\end{figure}

This counter-intuitive coupling scheme closely resembles the Stimulated Raman Adiabatic Passage (STIRAP) process in a three level atom.  Our system of three coupled harmonic resonators can be formally mapped onto such a three state system~\cite{Wang2011}.  The energy levels for a three state system are shown in Fig.~\ref{ThreeLevelAtom}.  If we identify the microwave cavity with state $\ket{1}$, the optical cavity with state $\ket{2}$, and the mechanical cavity with state $\ket{3}$, then, we can get population transfer from state $\ket{1}$ to $\ket{2}$ via the normal STIRAP process~\cite{stirap}, which leaves virtually no population in state $\ket{3}$.  In the STIRAP process, the Stokes coupling (the coherent coupling between states $\ket{2}$ and $\ket{3}$) is turned on first which splits the energy levels for state $\ket{3}$.  Then, the pump coupling (the coherent coupling between states $\ket{1}$ and $\ket{3}$) is turned on and the population in state $\ket{1}$ is seen to be transferred to state $\ket{2}$ without ever having any significant population in state $\ket{3}$ because of the interference between the pathways corresponding to transversing each of the two split energy levels.

\begin{figure}
  \includegraphics[width=85mm]{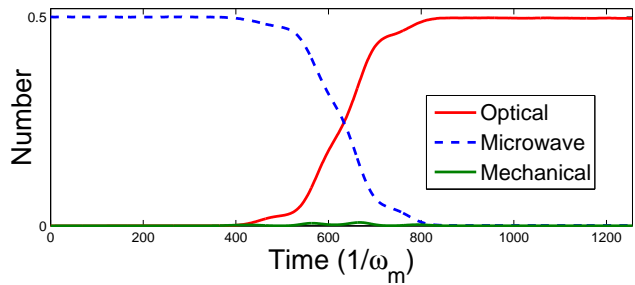}
  \caption{ (Color online) The number of photons or phonons in each resonator for the STIRAP-like coupling pulses.  The pulse area is $10 \pi$ and the peak separation is -14\%.}
  \label{EOM_AdiabaticNumber}
\end{figure}

For illustration purposes, we examine the populations of the three resonators through out this scheme shown in Fig.~\ref{EOM_AdiabaticNumber}.  Just as we would expect for a STIRAP-like process, the state adiabatically transfers from the microwave to the optical resonator while minimally populating the mechanical resonator.  Thus, this scheme is much more robust against decay and thermal noise as well as imperfections in pulse area.  However, adiabaticity generally requires longer times.  The simultaneous pulse scheme will transduce the state much quicker, but the pulses must be precisely generated.  So, there is a trade-off between the transduction time and the strigentness of pulse preparation and also decay during the simultaneous pulse width. 
 
A representative counter-intuitive partially overlapping coupling pulse configuration is shown in Fig.~\ref{TransferPulses}A by the black stars. For this case, the pulse area is $2.4 \pi$, the peak separation is $-15 \%$, and $\bar{n}=3$.

%%%%%%%%%%%%%%%%%%%%%%%%%%%%%%%%%%%%%%%%%%%%%%%%%%%%%%%%%%%%%%%%%%%%%%%%%%%%%%%%
\section{Conclusion}

Quantum state memory and transduction is possible for a broad range of experimentally achievable parameters in driven cavity optomechanics.  Many different types of states can be stored in the mechanical resonator and retrieved with a high fidelity in the range of experimentally achievable mechanical quality factors.  At the current experimental Q values of about $Q_m=360,000$~\cite{Teufel2011:couplingNumbers} the quantum state can be stored in the mechanical resonator for longer than ${\Delta}T=160 (units of 1/\omega_m)$ at low temperatures.  As the experiments improve the coupling strength, the time the state can be stored without significant degradation will increase.  If the mechanical mode is cooled before the swapping pulses are applied, then the storage time will also increase.

We have shown several procedures to accomplish transduction of quantum states.  High fidelity transfer is possible for Rabi type pulses of varying widths and separations even for very low Q mechanics.  We have shown that over 95\% transduction fidelity can be achieved for $Q_m>4525$ for $\Omega_m=0.1\omega_m$ and $\bar{n}=3$ for the simultaneous pulse scheme.  This scheme is quicker and more robust against thermal noise and decay than the more common separated pulse scheme.  Higher bath temperatures will require a shorter simultaneous pulse with stronger coupling to achieve high fidelities.

Counter-intuitively ordered adiabatic pulses can also be used to transfer the quantum state through a STIRAP like scheme with high fidelity.  This scheme is robust against thermal noise and decay and imperfections in pulse preparation and overlap, but requires longer transduction times.  Higher bath temperatures will also require longer transduction times to maintain the adiabaticity.

%%%%%%%%%%%%%%%%%%%%%%%%%%%%%%%%%%%%%%%%%%%%%%%%%%%%%%%%%%%%%%%%%%%%%%%%%%%%%%%%
\begin{acknowledgments}
% We would like to thank ----- for helpful discussions. 
This work was supported by the National Science Foundation.
\end{acknowledgments}

%%%%%%%%%%%%%%%%%%%%%%%%%%%%%%%%%%%%%%%%%%%%%%%%%%%%%%%%%%%%%%%%%%%%%%%%%%%%%%%%
\bibliography{QIStorageTransfer}

%%%%%%%%%%%%%%%%%%%%%%%%%%%%%%%%%%%%%%%%%%%%%%%%%%%%%%%%%%%%%%%%%%%%%%%%%%%%%%%%
\end{document}